\documentclass[sigconf,nonacm]{acmart}
\usepackage{tabularx}
\usepackage{algorithm,algorithmic}
\usepackage{placeins}
\usepackage{fancyvrb}
\definecolor{tbdcolor}{HTML}{B03A2E}
\definecolor{soft}{HTML}{F4F4F4}
\definecolor{trbg}{HTML}{FFF5F5}
\definecolor{trbd}{HTML}{E11D48}
\definecolor{trtx}{HTML}{9F1239}
\definecolor{grbg}{HTML}{EFF6FF}
\definecolor{grbd}{HTML}{2563EB}
\definecolor{grtx}{HTML}{1E3A8A}
\definecolor{revblue}{HTML}{1D4ED8}

\renewcommand{\arraystretch}{1.15}

\title{Shuttle: Dependency-Attributed Workflow Evolution for Smart Contract Auditing}
\author{Yue Xue}
\affiliation{\institution{CertiK}\city{Singapore}\country{Singapore}}
\email{nerbonic@gmail.com}

\author{Yi Rong}
\affiliation{\institution{Columbia University}\city{New York}\country{USA}}
\email{rong@cs.columbia.edu}

\author{Ziqiao Kong}
\affiliation{\institution{Nanyang Technological University}\city{Singapore}\country{Singapore}}
\email{ziqiao001@e.ntu.edu.sg}

\author{Ronghui Gu}
\authornote{Corresponding author.}
\affiliation{\institution{Columbia University}\city{New York}\country{USA}}
\email{ronghui.gu@columbia.edu}

\author{Dan She}
\affiliation{\institution{CertiK}\city{New York City}\country{USA}}
\email{dan.she@certik.com}

\author{Yuannan Yang}
\affiliation{\institution{CertiK}\city{New York City}\country{USA}}
\email{yuannan.yang@certik.com}

\author{Zhiju Yang}
\affiliation{\institution{Seattle University}\city{Seattle}\country{USA}}
\email{zhijuyang@seattleu.edu}

\author{Yuqiang Sun}
\affiliation{\institution{Nanyang Technological University}\city{Singapore}\country{Singapore}}
\email{yuqiang.sun@ntu.edu.sg}

\begin{document}
\begin{abstract}
In LLM-based smart contract auditing, detection quality depends on both model capability and workflow design. With stronger coding agents, models can now revise auditing workflows from execution feedback. Yet unconstrained revisions often fail to produce stable gains. Cross-function vulnerabilities in real contracts require multi-step state tracking and boundary reasoning that coarse pipelines miss. End-to-end outcomes rarely show which step lost a clue, so direct rewrites regress existing detections or discard valid edits under coarse aggregate scoring. Repeatedly iterating on a small set of projects also introduces case-specific heuristics that bloat prompts and hurt cross-project generalization. We present Shuttle, a constrained workflow-evolution system, together with LucidScan, the smart contract auditing framework it evolves. Shuttle centers on \textbf{provenance-graph attribution}. It expands each LucidScan run into an instance-level dependency graph, uses forward search to locate the node where a candidate clue was lost, and traces the source nodes of existing hits backward into a protected set. After a merged batch is evaluated, Shuttle removes edits that drag down performance directly through a node-level provenance ledger, falling back to cache-reusing local comparison re-runs only when coupled edits overlap on the same node in early rounds. For localized bottlenecks, Shuttle generates candidate edits under task decomposition, progressive refinement, and generality blind review, then advances the retained workflow. On ten training projects, Shuttle evolves LucidScan from baseline $G_0$ (28.7\% per-project average C/H/M recall) to $G_8$ (82.1\%). On thirteen validation projects not used in $M_{\mathrm{full}}$ evolution, under the same model, per-project average recall rises from 28.6\% at $G_0$ to 70.3\% at $G_8$, above Pashov and Codex Security on the same split. In an anonymized live public audit contest with 896 researchers concluded in November 2026, the evolved scanner tied for \#1 overall in High-severity recall (6/8, 75.0\%), ranked \#2 overall in scored H/M clusters (34/72, 47.2\%) with 92.9\% submission validity (78/84), and finished \#3 overall.

\end{abstract}
\maketitle
\section{Introduction}\label{sec:1}
LLM-based smart contract auditing depends on both base model capability and workflow design: how inspection scope is organized, how cross-function state interactions are tracked, and how boundary conditions are verified ~\cite{sun2024gptscan,liu2025propertygpt}. Manually tuning fixed workflows for real-world protocols is costly and often misses interaction patterns in new designs. Stronger coding and reasoning models have therefore enabled feedback-driven workflow evolution, moving from general agent synthesis ~\cite{hu2025adas,zhang2025dgm} into security auditing in systems such as AgentFlow ~\cite{liu2026agentflow}, EvoHunt ~\cite{wang2026evohunt}, and MetaVul ~\cite{ceka2026metavul}.

\begin{figure*}[t]
\centering
\includegraphics[width=\textwidth]{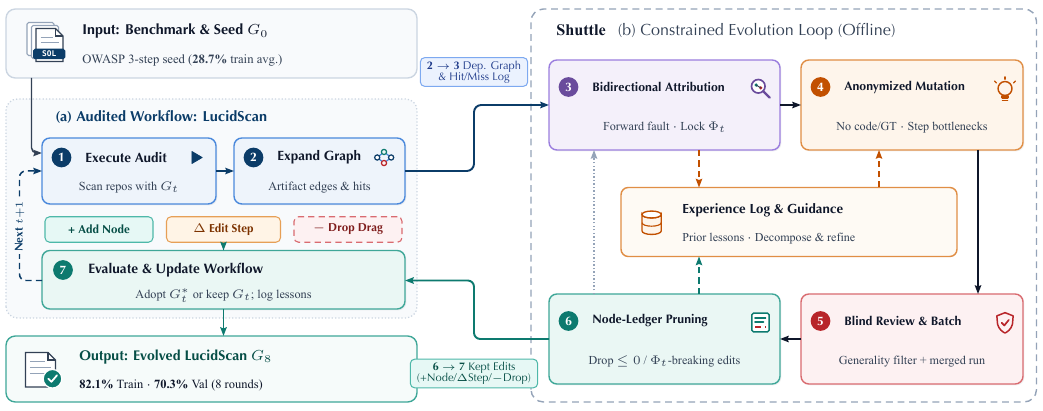}
\caption{Overview of how Shuttle evolves LucidScan's workflow from execution dependency records (read sequentially across Steps 1–7). Panel (a) executes the audit and expands the dependency graph (Steps 1–2); Panel (b) performs bidirectional graph attribution, anonymized mutation, generality blind review, and node-ledger pruning while the experience log synthesizes round outcomes and refinement guidance (Steps 3–6); Step 7 evaluates the retained workflow and updates LucidScan from $G_0$ to $G_8$ over eight rounds.}
\label{fig:1}
\end{figure*}

Applying workflow evolution to project-level contract auditing raises two problems. \textbf{First, end-to-end audit outcomes and flat execution logs neither locate which step of a multi-step pipeline lost a clue ~\cite{zhang2025whowhen,zhu2025agentdebug,ma2025dover} nor separate helpful edits from harmful ones in a merged round.} A final report lists only missed root causes. Once a pipeline branches in parallel, flat logs obscure artifact read-write dependencies and per-node outputs. Direct rewrites then regress existing hits, and when valid and flawed edits are tested together, coarse aggregate scores hide positive gains and cause valid edits to be discarded as regressions (\S\ref{sec:6.4}). \textbf{Second, iterating over a limited set of training projects invites case-specific heuristics and prompt bloat.} Evolving workflows on small task sets readily overfits project-specific patterns ~\cite{rrsi}, while stacking rules into long prompts distracts the model and degrades reasoning ~\cite{liu2024lost}.

To address both problems, we present \textbf{Shuttle}, a constrained workflow-evolution system, together with \textbf{LucidScan}, the smart contract auditing framework it evolves (Figure~\ref{fig:1}). Following the public OWASP Code Review Guide ~\cite{owasp2017codereview}, we initialize evolution with a three-step workflow $G_0$ (module decomposition, threat modeling, and reasoning). LucidScan runs the auditing workflow, while Shuttle keeps its attribution and generality rules fixed and records each round's attribution and outcomes in an \textbf{experience log}.

\textbf{First, Shuttle locates bottlenecks and filters edits through provenance-graph attribution.} Under the \textbf{Loom} runtime, each component declares the artifacts it reads and writes, recording task nodes and their read-write artifacts as a \textbf{dependency graph}. Provenance-graph attribution performs three operations on this graph. Forward search locates the node where a candidate clue was lost. Backward tracing identifies the source nodes of existing hits to form a \textbf{protected set}. After a merged batch is evaluated, the \textbf{node provenance ledger} directly removes edits that drag down performance, falling back to cache-reusing local comparison re-runs only when early-round edits overlap on the same node.

\textbf{Second, Shuttle generates transferable edits under generality constraints.} The proposer sees only stage-level process bottlenecks, with no project source code or ground-truth vulnerabilities, and revises the workflow using task decomposition and progressive refinement ~\cite{khot2023decomposed,madaan2023selfrefine}. Before evaluation, every candidate edit passes \textbf{generality blind review}, which rejects rules carrying project- or case-specific details.

With this loop, Shuttle evolves $G_0$ over eight rounds into the final workflow $G_8$. On thirteen validation projects not used in $M_{\mathrm{full}}$ evolution, per-project average C/H/M recall under the same model rises from 28.6\% at $G_0$ to 70.3\% at $G_8$, above Pashov and Codex Security. Section 6 answers four research questions:

\begin{itemize}
\item \textbf{RQ1 (Detection, Generalization, and Scan Cost):} How much does multi-round evolution improve recall on the training and validation sets, how does $G_8$'s recall compare with external tools, and what does one final scan cost?
\item \textbf{RQ2 (Weakly Constrained Evolution Lessons):} What cross-project divergence and token inflation appear in our early weakly constrained pilot ($M_{\mathrm{weak}}$) without generality blind review?
\item \textbf{RQ3 (Core Mechanisms and Local Pruning):} How do the full configuration and simplified ablations perform in the early-round window, and how do the node-level pruning decisions on $G_3$ and $G_7$ hold up in re-runs?
\item \textbf{RQ4 (Real-World Audit Contest):} On a live real-world audit contest, how many organizer-confirmed vulnerabilities does the evolved workflow report, and how does it rank in severity recall and overall standing?
\end{itemize}

We make two main contributions:

\begin{enumerate}
\item \textbf{The constrained workflow-evolution system Shuttle.} Built around \textbf{provenance-graph attribution} on an instance-level run dependency graph, Shuttle locates missed-clue nodes via forward search and traces stable-hit sources into a protected set via backward tracing. Combined with generality blind review and node-ledger pruning (plus cache-reusing local re-runs when edits couple), it removes negative or stability-harming edits while avoiding blind rewrites, mistaken rollbacks of valid edits, and case-specific overfitting.
\item \textbf{The evolved auditing framework LucidScan and its evaluation.} We report $G_8$'s validation recall (70.3\% vs. 28.6\% for $G_0$ under the same model), external-tool comparisons, scan cost (USD 4.92 per validation project per run), early-round ablations, $G_3$/$G_7$ pruning re-runs, and live results from an anonymized public audit contest concluded in November 2026.
\end{enumerate}

Section 2 motivates our design. Sections 3–5 present the problem definition, method, and implementation. Section 6 answers RQ1–RQ4, and Sections 7–9 cover related work, limitations, and conclusions.

\section{Motivation}\label{sec:2}
\subsection{Four local breakpoints behind a single miss}\label{sec:2.1}

To illustrate why diagnosing failures in multi-step auditing workflows requires intermediate visibility, consider the \texttt{ShareVault} snippet below, where \texttt{deposit} mints shares \texttt{s} and \texttt{withdraw} burns \texttt{s} before transferring funds via \texttt{call}:

\par\smallskip\noindent\begin{minipage}{\columnwidth}
\begin{Verbatim}[frame=single,rulecolor=\color{black!30},framesep=4pt,fontsize=\scriptsize]
contract ShareVault {
  uint256 public totalShares; uint256 public totalAssets;
  function deposit(uint256 assets) external {
    uint256 s = assets * totalShares / totalAssets;
    _mint(msg.sender, s); totalShares += s; totalAssets += assets;
  }
  function withdraw(uint256 assets) external {
    uint256 s = assets * totalShares / totalAssets;
    _burn(msg.sender, s); totalShares -= s; totalAssets -= assets;
    payable(msg.sender).call{value: assets}("");
  }
}
\end{Verbatim}
\end{minipage}\par\smallskip

This contract contains two levels of defects:

\begin{enumerate}
\item \textbf{Local surface flaws:} \texttt{withdraw} leaves \texttt{call} unchecked and \texttt{deposit} lacks a \texttt{totalAssets == 0} guard. Both can be caught by inspecting one function in isolation.
\item \textbf{Zero-share free-withdrawal bug:} Both functions truncate \texttt{s = assets * totalShares / totalAssets} downward. When prior yield or donation leaves \texttt{totalAssets > totalShares} (e.g., \texttt{100 > 10}), a small withdrawal (\texttt{assets = 1}) truncates \texttt{s} to \texttt{0}. If \texttt{\_burn} does not revert on \texttt{s == 0}, a caller can extract surplus assets without burning shares (e.g., 90 calls to \texttt{withdraw(1)} take 90 units of surplus until \texttt{totalAssets == totalShares == 10}).
\end{enumerate}

Detecting this bug follows a four-step path: \textbf{place functions sharing ledger state in one review scope → propose the share-to-asset conversion hypothesis → pass the formula and the \texttt{s = 0} boundary downstream → substitute \texttt{s = 0} during reasoning}. If an audit reports only the \texttt{call} flaw and misses the zero-share withdrawal, the final output cannot reveal where the pipeline failed. Externally identical misses can come from \textbf{four distinct local breakpoints}:

\begin{itemize}
\item \textbf{Missing Scope:} \texttt{withdraw} and other functions sharing the same ledger state are placed in separate units, so the model never inspects the full state-update path.
\item \textbf{Absent Candidate:} The relevant functions are in scope, but analysis stops at surface call flaws and never proposes share truncation.
\item \textbf{Lost Handoff:} Upstream notes the formula and the \texttt{s = 0} boundary, yet compresses them into a vague “precision loss” summary, leaving downstream without concrete conditions.
\item \textbf{Flawed Judgment:} Downstream receives the \texttt{s = 0} clue but dismisses it by intuition (“rounding down favors the vault”) without checking \texttt{\_burn(msg.sender, 0)} with concrete values.
\end{itemize}

Coarse steps collapse all four failures into one macro-step. Attribution can then point only to the whole step, forcing the proposer to stack ad-hoc rules into the same prompt. Those rules miss the actual breakpoint, disturb working detection paths, and lengthen the prompt. Splitting the pipeline into explicit artifact-passing steps lets the dependency graph separate where a clue was lost while tracing the source nodes of existing hits.

\subsection{Overfitting, prompt bloat, and multi-edit interference}\label{sec:2.2}

Repeatedly editing a workflow on a limited training set creates three additional risks. First, \textbf{case-specific overfitting}: if the proposer sees \texttt{ShareVault} directly, it tends to write contract-specific rules such as checking whether \texttt{\_burn} truncates to zero. Second, \textbf{prompt bloat}: stacking ad-hoc rules into one step lengthens the context and raises scan token cost (the weakly constrained pilot $M_{\mathrm{weak}}$ in \S\ref{sec:6.3} shows a 26.6× inflation in per-project scan tokens alongside sharp validation divergence). Third, \textbf{multi-edit interference}: when several edits are merged in one round, an aggregate score gain can hide negative or redundant branches (\S\ref{sec:6.4}). Shuttle addresses these risks with role-level information isolation, generality blind review, node-ledger local pruning, and an experience log, replacing rule stacking with task decomposition and progressive refinement.

\section{Problem Definition}\label{sec:3}
\subsection{Evolution target and role isolation}\label{sec:3.1}

An audit workflow $G$ coordinates multi-step LLM calls and deterministic scripts over a project's source code and documentation. We initialize $G_0$ as a three-step pipeline following the OWASP Code Review Guide ~\cite{owasp2017codereview}. Given $G_0$ and training split $\mathcal{D}_{\mathrm{train}}$, Shuttle searches legal workflow revisions $\Delta$ yielding $G' = \Delta(G)$ while keeping inter-step input/output dependencies valid and meeting the generality rules in \S\ref{sec:4.3}. Throughout evolution, model weights and outer rules such as graph attribution, role isolation, and generality blind review remain fixed. Only the workflow $G$ evolves, while the Shuttle experience log is updated across rounds to record each round's attribution and regression lessons.

To prevent the case-specific overfitting in \S\ref{sec:2.2}, benchmark projects are split at repository boundaries into training $\mathcal{D}_{\mathrm{train}}$ and validation $\mathcal{D}_{\mathrm{val}}$, and role inputs are isolated. Auditing and attribution read project source code and run records, whereas the proposer and blind reviewer never see project names, source code, or ground-truth vulnerabilities (\S\ref{sec:4.3}).

\subsection{Evaluation metrics}\label{sec:3.2}

An audit finding counts as a hit when its root cause matches a ground-truth (GT) vulnerability, at most once per GT per run. Over evaluation split $\mathcal{P}$ with $k$ independent runs per project, known issues $\mathcal{B}_p$ ($|\mathcal{B}_p|>0$), total issues $M=\sum_{p\in\mathcal{P}}|\mathcal{B}_p|$, and hit indicator $x_{b,r}(G)\in\{0,1\}$, we define single-project recall $R_p(G)$, \textbf{per-project average recall} $\bar{R}_{\mathrm{proj}}(G)$, and multi-run union coverage $U(G)$: \begin{gather*}R_p(G)=\frac{\sum_{b\in\mathcal{B}_p}\sum_{r=1}^{k}x_{b,r}(G)}{k|\mathcal{B}_p|}\\
\bar{R}_{\mathrm{proj}}(G)=\frac{1}{|\mathcal{P}|}\sum_{p\in\mathcal{P}} R_p(G)\\
U(G)=\frac{\sum_{p\in\mathcal{P}}\sum_{b\in\mathcal{B}_p}\mathbb{I}[\sum_{r=1}^{k}x_{b,r}(G)>0]}{M}\end{gather*} We use \textbf{per-project average recall} $\bar{R}_{\mathrm{proj}}(G)$ as our primary metric, averaging each project's $k$-run recall equally across projects so issue-heavy repositories do not dominate. We also report multi-run union coverage $U(G)$ as a secondary indicator, together with scan cost.

\section{Method}\label{sec:4}
Each Shuttle round follows Figure~\ref{fig:1} and Algorithm 2 in four steps: run the current workflow and expand its \textbf{dependency graph} (\S\ref{sec:4.1}); run \textbf{bidirectional attribution} on the graph (\S\ref{sec:4.2}); generate edits under guidance and \textbf{generality blind review} (\S\ref{sec:4.3}); and, after merged evaluation, prune by the \textbf{node provenance ledger} and select a version (\S\ref{sec:4.4}). The dependency graph, bidirectional attribution, and ledger pruning share one graph and are together called \textbf{provenance-graph attribution}.

\subsection{Run dependency graph: what one run records}\label{sec:4.1}

The \textbf{Loom} runtime represents each step as a declarative component $c=(\mathrm{In}_c,\mathrm{Out}_c,f_c,\pi_c)$: the artifacts it reads, the artifacts it writes, its operation (an LLM call or a script), and its execution policy (per unit, parallel, or aggregate). When a task instance finishes, the executor leaves three kinds of records, and attribution reads only these (Figure~\ref{fig:2}A):

\begin{enumerate}
\item \textbf{Artifacts and provenance records.} Outputs are stored by content hash together with the producing instance, the hashes of upstream artifacts read, and the version of prompts and other assets. If instance $v$ reads an artifact of instance $u$, we add an edge $u\to v$; the run dependency graph $G^{\mathrm{run}}(G,p,r)=(V,E)$ is assembled directly from these records, with task instances as nodes.
\item \textbf{Candidate IDs.} Candidates carry hierarchical module–candidate–report IDs that downstream nodes keep verbatim, so the same candidate can be located exactly in every node's artifacts. Nodes added during evolution put a distinguishable suffix into their IDs, from which each final report records the nodes it passed through.
\item \textbf{Execution records and transcripts.} Reasoning and similar nodes attach a lightweight execution record (files read, commands run) that attribution may read; full transcripts are archived only and never read.
\end{enumerate}

Shuttle writes a \textbf{merge manifest} recording the node set $V(e)$ modified or added by each edit $e$. Each edit targets a single local bottleneck, and mutually exclusive edits to the same node go into separate batches.

\subsection{Bidirectional attribution: misses search downstream for a breakpoint, hits search upstream for a source}\label{sec:4.2}

\begin{figure*}[t]
\centering
\includegraphics[width=\textwidth]{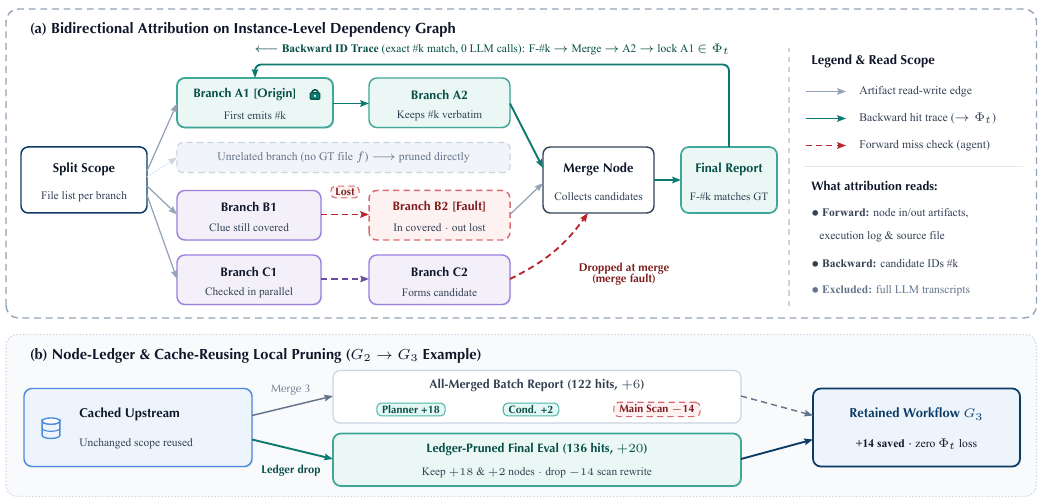}
\caption{Provenance-graph attribution and local pruning on the instance-level dependency graph. (a) Each artifact carries a provenance record (producing node instance, hashes of upstream artifacts read, version) and candidate IDs, while full LLM transcripts are archived rather than read. For a miss (purple and crimson dashed lines), a program prunes unrelated branches, an agent inspects each node's input and output artifacts to judge coverage, and the search advances until an "input covered, output lost" fault node or merge fault. For a hit (teal solid lines), a program traces candidate IDs \texttt{\#k} in parent artifacts back to the earliest origin node and locks it into the protected set $\Phi_t$ with zero LLM calls. (b) Node provenance ledger: intersecting the nodes each hit traversed with the nodes each edit modifies isolates per-node net credit and drops negative-drag edits ($G_2 \to G_3$: 122 $\to$ 136 hits).}
\label{fig:2}
\end{figure*}

Attribution reads $G^{\mathrm{run}}$, the GTs, the match outcomes, and the source snapshot (Figure~\ref{fig:2}A, Algorithm 1). Each GT is routed by its hit count $h$ over $k=3$ runs: $h\ge 2$ goes to backward tracing, all others to forward localization.

\begin{algorithm}[!t]
\footnotesize
\caption{Bidirectional graph search and attribution}\label{alg:1}
\begin{algorithmic}[1]
\REQUIRE Dependency graph $G^{\mathrm{run}}$, target GT $b$, hit count $h\in\{0,\dots,k\}$ ($k=3$), source snapshot
\ENSURE Coverage-flip breakpoint(s), isolated unknowns, or protected origin set $\mathcal{S}_t(b)$
\STATE Prune $G^{\mathrm{run}}$ to the subgraph $G_b$ relevant to target $b$ \COMMENT{keep GT-relevant units and index edges}
\IF{$h \le 1$}
\STATE Initialize worklist $Q$ over $G_b$; inspect node input/output coverage via coding agent $\to \{\text{covered},\text{partial},\text{uncovered},\text{unknown}\}$ \COMMENT{miss/unstable: four-state worklist search}
\WHILE{worklist $Q \neq \emptyset$ (within query budget)}
\STATE Pop unvisited $u \in Q$ and mark visited; if $\text{input}=\text{covered}$ and $\text{output}=\text{uncovered}$, record breakpoint and skip expansion \COMMENT{require cited input and output evidence}
\IF{$\text{output}=\text{covered}$} \STATE enqueue children into $Q$
\ELSIF{$\text{output}=\text{uncovered}$} \STATE enqueue relevant parent nodes into $Q$
\ELSIF{$\text{output}=\text{partial}$} \STATE enqueue both children and relevant parent nodes into $Q$
\ELSE \STATE put node in the unknown set \COMMENT{$\text{output}=\text{unknown}$ or provenance incomplete}
\ENDIF
\ENDWHILE
\STATE Flag merge nodes dropping incoming candidates; inspect all parallel generators
\STATE \textbf{return} coverage-flip breakpoint(s) or isolated unknown records
\ELSE
\STATE Locate terminal producer(s) by matched finding ID, then follow dependency edges backward, matching the same candidate ID in parent artifacts, to the earliest node(s) $\mathcal{S}_t(b)$ where the candidate appears (no model call) \COMMENT{$h \ge 2$: backward tracing}
\STATE \textbf{return} $(b, \mathcal{S}_t(b))$ for protected set $\Phi_t$
\ENDIF
\end{algorithmic}
\end{algorithm}

\textbf{Forward localization (misses): find the node where the clue goes from present to absent.}

\begin{itemize}
\item \textbf{Prune (program).} Take the files named by the GT (if none, infer them from the function names the GT mentions) and check only branches whose file list contains these files.
\item \textbf{Judge one node (agent).} The agent receives one GT and one node at a time: its input and output artifacts, the source, and the node's responsibility. It labels whether the input and the output each cover the target mechanism (covered, partial, uncovered, unknown), citing artifact text. A reasoning node's input covers only if it already contains the candidate.
\item \textbf{Advance (program).} Covered output moves downstream; covered input with uncovered output is the breakpoint; uncovered input moves upstream; \texttt{unknown} or incomplete records stop the search. If a merge node drops a candidate formed upstream, the breakpoint is that merge node.
\item \textbf{Result.} Each breakpoint carries a failure type (\S\ref{sec:2.1}: scope omission, candidate not formed, handoff loss, reasoning misjudgment, plus "judged correctly but not collected"), aggregated by step and type for the proposer (\S\ref{sec:4.3}).
\end{itemize}

\textbf{Backward tracing (stable hits): follow IDs to the earliest producer, without a model call.}

\begin{itemize}
\item \textbf{Match.} Starting from the matched report's ID, look for the same ID or verbatim identical text in parent outputs and continue upstream while found; only fields are compared, with no semantic judgment.
\item \textbf{Follow only parents containing the ID.} A node may read several parents; a parent whose output lacks the ID never carried this candidate.
\item \textbf{Protected set.} The earliest node reached is the source; the sources of all stable hits form the \textbf{protected set} $\Phi_t$, and edits touching $\Phi_t$ must pass the regression check in \S\ref{sec:4.4}.
\end{itemize}

\subsection{Workflow revision, experience log, and generality blind review}\label{sec:4.3}

Shuttle separates \textbf{where a step failed} from \textbf{how to revise it} with three mechanisms (Figure~\ref{fig:1} Steps 4–5):

\begin{enumerate}
\item \textbf{Information isolation and experience log.} Attribution results are aggregated by step and failure type, stripping project names, identifiers, source, and GT details; the proposer sees only step-level bottlenecks and past successes and failures from the experience log.
\item \textbf{Task decomposition and progressive refinement.} The proposer improves the workflow by splitting steps and adding intermediate evidence rather than stacking bug checklists into one node; edits fall into a text layer and a graph layer (Table~\ref{tab:1}).
\item \textbf{Generality blind review.} The hard invariant is that \textbf{evolution may only revise \emph{how to audit}, never \emph{what bug to find}}. Independent blind review drops edits carrying project- or case-specific traits; surviving conflict-free edits $E_t$ form $G_t^{\mathrm{cand}}=G_t\oplus E_t$ for evaluation on $\mathcal{D}_{\mathrm{train}}$.
\end{enumerate}

\begin{table}[!tb]
\centering\small
\caption{Two layers of workflow modification actions in Shuttle (text/field edits and graph-structure edits).}\label{tab:4.0}\label{tab:1}
\begin{tabularx}{\columnwidth}{@{}l>{\raggedright\arraybackslash}X>{\raggedright\arraybackslash}X@{}}
\toprule
Layer & Basic Modification Actions & Common Combinations \& Role \\
\midrule
Text & \texttt{edit fragment}, \texttt{add/delete fragment}, \texttt{edit prompt}, \texttt{edit schema field} & \texttt{add refinement}: smallest-magnitude prompt and output-field edits \\
Graph Topology & \texttt{add node}, \texttt{delete node}, \texttt{rewire edge}, \texttt{edit execution policy} & \texttt{split step}, \texttt{add parallel lane}, \texttt{add verification}, \texttt{change granularity}, \texttt{merge steps} \\
\bottomrule
\end{tabularx}
\end{table}

\subsection{Local pruning and selection}\label{sec:4.4}

When a round merges several edits $e\in E_t$, a higher total can still hide useless or harmful ones. The \textbf{node provenance ledger} intersects the nodes each final report passed through with each edit's $V(e)$ (Figure~\ref{fig:2}B), assigning hits and choosing how to remove edits as in Table~\ref{tab:8}. The ledger gives ownership, not causation, so it is used directly only when edits separate cleanly.

\begin{table}[!tb]
\centering\small
\caption{Hit assignment and edit removal rules of the node provenance ledger.}\label{tab:4.1}\label{tab:8}
\begin{tabularx}{\columnwidth}{@{}l>{\raggedright\arraybackslash}X>{\raggedright\arraybackslash}X@{}}
\toprule
Step & Case & Handling \\
\midrule
Assign & Hit intersects one edit & Exclusive to that edit; counts toward its net gain \\
Assign & Hit intersects several edits & Shared/unassigned; credited to no single edit \\
Assign & Hit intersects no edit & Baseline \\
Assign & Old hit lost in the new version & Look up the nodes of the old hit; each edit that modified one of them is charged one loss \\
Remove & Edits separable along dependency edges (e.g., $G_3,G_6,G_7,G_8$) & Drop edits with net gain $\le 0$ or $\Phi_t$ regressions directly \\
Remove & Edits share a node or are coupled (e.g., $G_0\to G_1$) & Freeze upstream caches and rerun $G_t^{\mathrm{cand}}$ vs. $G_t^{\mathrm{cand}}\setminus\{e\}$ locally \\
\bottomrule
\end{tabularx}
\end{table}

The pruned $G_t^*$ is evaluated on $\mathcal{D}_{\mathrm{train}}$ and judged on C/H/M recall, $\Phi_t$ stability, union coverage, and cost: if it passes, $G_{t+1}\leftarrow G_t^*$; otherwise $G_t$ is kept. Outcomes go into the experience log.

\begin{algorithm}[!t]
\small
\caption{One round of evolution}\label{alg:2}
\begin{algorithmic}[1]
\REQUIRE $G_t$, evaluated run records, modification actions, generality constraints, and experience log
\STATE Execute $G_t$ across $\mathcal{D}_{\mathrm{train}}$ (or reuse version-matched records from the prior round)
\STATE Expand dependency graph $G^{\mathrm{run}}(G_t,p,r)$, logging content hashes and match outcomes
\STATE Stratify GTs by $k$-run hit count $h\in\{0,\dots,k\}$ ($k=3$) and run bidirectional graph attribution on $G^{\mathrm{run}}$ via deterministic indexing plus per-GT local search to localize miss bottlenecks and lock $\Phi_t$
\STATE Aggregate stage-level process bottlenecks and generate candidate edits guided by task decomposition, progressive refinement, and the experience log
\STATE Apply generality blind review; merge conflict-free edits $G_t^{\mathrm{cand}}=G_t\oplus E_t$ and evaluate on $\mathcal{D}_{\mathrm{train}}$
\STATE Prune edits with net gain $\le 0$ or $\Phi_t$ regressions directly from the node provenance ledger (or freeze compatible upstream caches for paired leave-one-out reruns $G_t^{\mathrm{cand}}\setminus\{e\}$ when edits are coupled) to obtain $G_t^*$
\STATE Evaluate retained $G_t^*$ on $\mathcal{D}_{\mathrm{train}}$, decide acceptance ($G_{t+1}\leftarrow G_t^*$ or retain $G_t$), and update the experience log
\ENSURE $G_{t+1}$ and its evaluated records
\end{algorithmic}
\end{algorithm}

\section{Implementation}\label{sec:5}
\begin{figure*}[t]
\centering
\includegraphics[width=\textwidth]{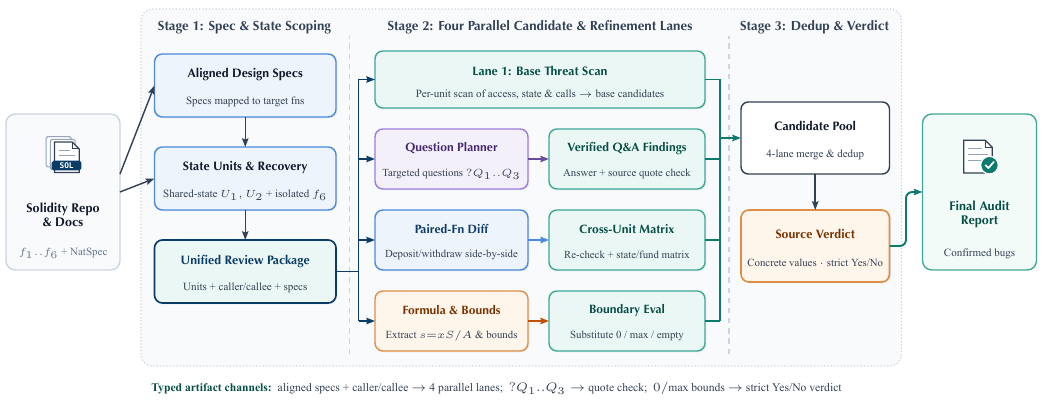}
\caption{Three-stage workflow architecture of LucidScan $G_8$ evolved from $G_0$'s three-step baseline. Each step of $G_0$ expands into a structured stage: Stage 1 extracts design specifications from documentation and partitions shared-state review units (recovering isolated functions); Stage 2 executes four parallel candidate-generation and progressive-refinement lanes; Stage 3 deduplicates candidates, verifies substituted boundary values per candidate, and emits the final audit report.}
\label{fig:3}
\end{figure*}

The system has three layers: \textbf{Loom}, a vulnerability-agnostic multi-agent runtime that parses read-write interfaces, schedules concurrent calls, and persists artifacts; \textbf{LucidScan}, the auditing workflow; and \textbf{Shuttle}, which runs attribution, proposal, blind review, and pruning offline only, so the finalized $G_8$ runs directly on Loom. Each of $G_0$'s three steps becomes one layer of $G_8$ (Figure~\ref{fig:3}):

\begin{itemize}
\item \textbf{Layer 1 (decomposition).} Extract design expectations from docs and comments and align them to functions; partition state-sharing review units and bring ungrouped functions back into scope.
\item \textbf{Layer 2 (modeling).} Run four parallel lanes: main threat modeling; supplementary inquiry (plan questions, answer independently, check alignment); unit comparison with a relation matrix; and condition derivation with boundary evaluation.
\item \textbf{Layer 3 (reasoning).} Merge and deduplicate candidates from all lanes, then judge each with concrete value substitution and source evidence.
\end{itemize}

\section{Evaluation}\label{sec:6}
This section answers the four research questions of \S\ref{sec:1}. $M_{\mathrm{full}}$ is the full \S\ref{sec:4} method; $M_{\mathrm{weak}}$ is the early weakly constrained pilot; $M_{\mathrm{guide}}$, $M_{\mathrm{trace}}$, and $M_{\mathrm{miss}}$ are ablations.

\subsection{Dataset and evaluation protocol}\label{sec:6.1}

The early $M_{\mathrm{weak}}$ pilot (\S\ref{sec:6.3}) evolved on a chronologically collected project set and reached high recall on token contracts but dropped sharply on lending and vault protocols. $M_{\mathrm{full}}$ therefore uses a rebuilt benchmark of 23 real projects from 2026 public audit contests and confidential internal audits (project names are anonymized as \texttt{T1}–\texttt{T10} for training and \texttt{V1}–\texttt{V13} for validation); GTs are the issues accepted by each project team in its final audit report. Following PropertyGPT ~\cite{liu2025propertygpt}, we cluster source embeddings of 311 deduplicated contracts with Affinity Propagation into 10 protocol clusters (Appendix Figure~\ref{fig:4}) and split by repository into 10 training projects (covering 8 clusters) and 13 validation projects (Table~\ref{tab:2}); the remaining settings are in Table~\ref{tab:9}.

\begin{table}[!tb]
\centering\small
\caption{Protocol-type clustering of the 23 benchmark projects by source-code similarity (10 clusters) and the 10/13 train–validation split (203 training GTs with 79 C/H/M; 254 validation GTs with 114 C/H/M).}\label{tab:6.0}\label{tab:2}
\begin{tabularx}{\columnwidth}{@{}>{\raggedright\arraybackslash}X>{\raggedright\arraybackslash}X@{}}
\toprule
Cluster \& Category & Projects (Train / Val) \\
\midrule
\texttt{C0} Token Sale \& Vesting & T2 / — \\
\texttt{C1} Staking Proxy & — / V2 \\
\texttt{C2} NFT Lending \& Liquidation & T6 / V5 \\
\texttt{C3} LST \& Strategy Vaults & — / V6, V12 \\
\texttt{C4} NFT Market \& Bonding & T5 / V7 \\
\texttt{C5} Staking \& Distribution & T3, T7 / V9 \\
\texttt{C6} DEX, Gauges \& Routers & T1, T4 / V1, V11, V13 \\
\texttt{C7} Mining \& Token & T8 / — \\
\texttt{C8} Meme Token \& Distributor & T9 / — \\
\texttt{C9} AMM Hooks \& Leverage & T10 / V3, V4, V8, V10 \\
\textbf{Total} & \textbf{10 Train (8/10) / 13 Val} \\
\bottomrule
\end{tabularx}
\end{table}

\begin{table}[!tb]
\centering\small
\caption{Experimental settings: GT scale, runs, optimization target, selection basis, and models per role.}\label{tab:6.1}\label{tab:9}
\begin{tabularx}{\columnwidth}{@{}>{\raggedright\arraybackslash}X>{\raggedright\arraybackslash}X@{}}
\toprule
Setting & Value \\
\midrule
GT scale & Train 203 (79 C/H/M); validation 254 (114 C/H/M) \\
Runs & Train: 3 per version; validation: 3 for $G_0$ and $G_8$, 2 for $G_6$ and $G_7$; external baselines: 1 with default settings \\
Optimization target & Attribution, proposal, and selection target C/H/M issues only \\
Version selection & Training results only; validation plays no part \\
Scanning model & \texttt{Luna high} (shared by $G_0$, $G_8$, and Pashov Luna) \\
Attribution \& proposal & \texttt{Sol medium}; proposer isolated from blind review \\
Root-cause judge & \texttt{Sol high}, tool-blinded ~\cite{wang2026evohunt}; a heterogeneous model agrees on 58 of 60 stratified samples ($\kappa=0.933$) \\
Runtime & Up to 8 concurrent agents, 40-minute timeout, 1 retry; separate cache per run \\
\bottomrule
\end{tabularx}
\end{table}

\subsection{RQ1: Detection Effectiveness, Generalization, and Scan Cost}\label{sec:6.2}

\textbf{Question.} How much does multi-round evolution improve LucidScan's recall and coverage, how does $G_8$'s cross-project recall compare with external baselines, and what is its per-scan API cost?

\begin{figure*}[t]
\centering
\includegraphics[width=\textwidth]{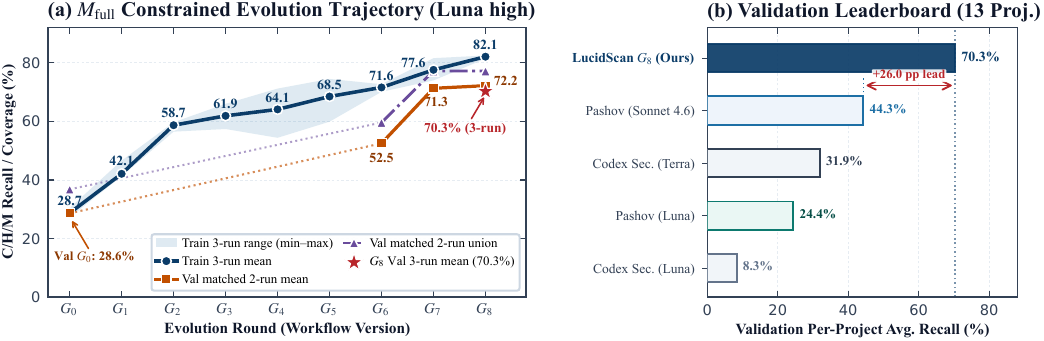}
\caption{(a) Constrained evolution trajectory of $M_{\mathrm{full}}$ ($G_0$–$G_8$) under Luna high (3-run mean and min–max band over 10 training projects; final $G_8$ detects 64/79 C/H/M GTs in the 3-run union), paired with the validation $G_0$ baseline (28.6\%, 3-run mean), matched 2-run validation means and unions for $G_6$–$G_8$, and the final $G_8$ 3-run validation mean of 70.3\% (star). (b) Validation per-project C/H/M recall comparison across all 13 unseen projects against single-run external baselines.}
\label{fig:5}
\end{figure*}

\textbf{Training.} Eight rounds raise C/H/M per-project average recall from 28.7\% to 82.1\%, with a 1.1 pp three-run spread at $G_8$ (Figure~\ref{fig:5}, Appendix Table~\ref{tab:6}).

\textbf{Validation.} With the same model, $G_8$ raises C/H/M per-project average recall from 28.6\% ($G_0$) to 70.3\% (11 projects improve, 2 tie), and three-run union coverage from 36.8\% to 78.9\%, above every external tool (Table~\ref{tab:3}). On the 5 projects evaluated by KnowDit ~\cite{kong2026knowdit}, counting all severities, $G_8$ reaches 61.9\% and KnowDit 24.9\%.

\begin{table*}[t]
\centering\small
\caption{Per-project C/H/M recall across all 13 validation benchmark projects ($G_0$ and $G_8$: 3-run mean; external baselines: default single pass; row-wise maximums bolded; $\dagger$ KnowDit evaluated on 5 validation projects under all-severity ground truth; its V6 run hit the USD 400 budget cap).}\label{tab:6.2}\label{tab:3}
\begin{tabularx}{\textwidth}{@{}l>{\raggedright\arraybackslash}X>{\raggedright\arraybackslash}X>{\raggedright\arraybackslash}X>{\raggedright\arraybackslash}X>{\raggedright\arraybackslash}X>{\raggedright\arraybackslash}X>{\raggedright\arraybackslash}X>{\raggedright\arraybackslash}X@{}}
\toprule
Project (C/H/M, All) & $G_0$ (C/H/M) & $M_{\mathrm{full}}$ $G_8$ (C/H/M) & Pashov (Son.) & Pashov (Ter.) & Pashov (Luna) & CS (Terra) & CS (Luna) & KnowDit \\
\midrule
V1 (5, 17) & 47\% & \textbf{80\%} & \textbf{80\%} & \textbf{80\%} & 0\% & 60\% & 0\% & — \\
V2 (3, 5) & 33\% & \textbf{67\%} & 33\% & \textbf{67\%} & \textbf{67\%} & 33\% & 0\% & — \\
V3 (1, 1) & 0\% & 0\% & 0\% & 0\% & 0\% & 0\% & 0\% & — \\
V4 (17, 31) & 20\% & \textbf{73\%} & 59\% & 41\% & 6\% & 29\% & 6\% & 48\% \\
V5 (37, 37) & 13\% & \textbf{67\%} & 49\% & 35\% & 16\% & 32\% & 5\% & 27\% \\
V6 (2, 8) & 17\% & \textbf{100\%} & 0\% & 0\% & 0\% & 0\% & 0\% & 13\% \\
V7 (3, 3) & 44\% & 89\% & \textbf{100\%} & \textbf{100\%} & \textbf{100\%} & 67\% & 33\% & — \\
V8 (7, 58) & 0\% & \textbf{38\%} & 14\% & 0\% & 14\% & 0\% & 0\% & 10\% \\
V9 (21, 46) & 41\% & \textbf{86\%} & 33\% & 24\% & 14\% & 29\% & 14\% & — \\
V10 (13, 19) & 8\% & \textbf{49\%} & 8\% & 8\% & 0\% & 15\% & 0\% & 26\% \\
V11 (2, 25) & \textbf{100\%} & \textbf{100\%} & \textbf{100\%} & \textbf{100\%} & \textbf{100\%} & 50\% & 50\% & — \\
V12 (2, 3) & 17\% & \textbf{100\%} & \textbf{100\%} & \textbf{100\%} & 0\% & \textbf{100\%} & 0\% & — \\
V13 (1, 1) & 33\% & \textbf{67\%} & 0\% & 0\% & 0\% & 0\% & 0\% & — \\
\textbf{Per-Proj Avg} & \textbf{28.6\%} & \textbf{70.3\%} & 44.3\% & 42.7\% & 24.4\% & 31.9\% & 8.3\% & 24.9\%$\dagger$ \\
\bottomrule
\end{tabularx}
\end{table*}

\textbf{Cost.} $G_8$ costs USD 4.92 in direct API spend per project per run (Appendix Table~\ref{tab:7}).

\textbf{Summary (RQ1).} With the same model, evolution lifts validation C/H/M recall from 28.6\% to 70.3\%, above all external tools.

\subsection{RQ2: Generalization lessons from weakly constrained evolution ($M_{\mathrm{weak}}$)}\label{sec:6.3}

\begin{figure}[!tb]
\centering
\includegraphics[width=\columnwidth]{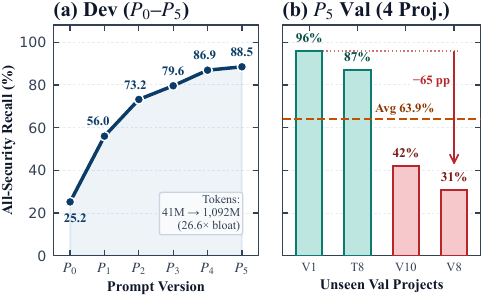}
\caption{Weakly constrained $M_{\mathrm{weak}}$ evaluation: (a) sequential development recall across $P_0$–$P_5$ and (b) $P_5$ transfer recall across four unseen validation protocols, illustrating sharp cross-protocol divergence.}
\label{fig:6}
\end{figure}

\textbf{Question.} In early weakly constrained evolution ($M_{\mathrm{weak}}$) without generality blind review, what cross-project divergence and token inflation emerge?

$M_{\mathrm{weak}}$ ran on a separate chronologically collected set of 10 development projects (226 security GTs) and 4 validation projects. Five rounds lift development recall from 25.2\% to 88.5\% while per-project scan tokens grow 26.6× (Figure~\ref{fig:6}, left). On validation, token protocols stay high (\texttt{V1} 96\%, \texttt{T8} 87\%) while \texttt{V10} (42\%) and \texttt{V8} (31\%) drop, for a 63.9\% per-project average (52.9\% pooled; Figure~\ref{fig:6}, right).

\textbf{Summary (RQ2).} Evolution without blind review climbs on the development set but brings 31\%–96\% cross-project divergence and 26.6× token inflation, which led $M_{\mathrm{full}}$ to add case-stripped handoffs, generality blind review, and a protocol-diverse split.

\subsection{RQ3: Core mechanism ablation, local pruning, and provenance comparison}\label{sec:6.4}

\textbf{Question.} In the early-round ablation window, how do the full configuration and the simplified ablations perform? How do the node-level pruning decisions on $G_3$ and $G_7$ fare in re-runs?

\begin{figure}[!tb]
\centering
\includegraphics[width=\columnwidth]{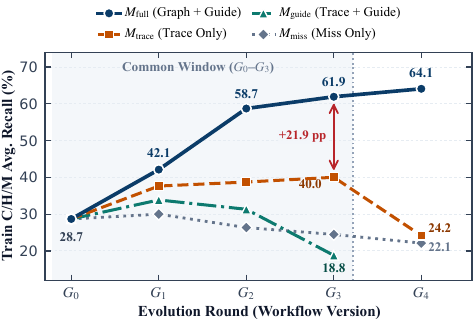}
\caption{Early-round C/H/M per-project average recall over $G_0$–$G_4$ (3-run mean per project; shaded: common $G_0$–$G_3$ window). The full configuration ($M_{\mathrm{full}}$) keeps rising, whereas the three simplified ablations all fall back.}
\label{fig:7}
\end{figure}

\textbf{(1) Early-round component ablation.} The four configurations share data, scanning, and judging models and remove components step by step from the full one (Table~\ref{tab:10}). The three graph-free configurations cannot use the ledger and advance from each round's merged output; whole-round rollback never occurred in eight rounds. The full configuration keeps rising while all three ablations fall back within the early window (Figure~\ref{fig:7}).

\begin{table*}[t]
\centering\small
\caption{Early-round ablation: composition of the four configurations and training C/H/M per-project average recall over $G_1$–$G_4$ ($G_0$ is 28.7\% for all; $M_{\mathrm{guide}}$ runs only through $G_3$).}\label{tab:6.3}\label{tab:10}
\begin{tabularx}{\textwidth}{@{}l>{\hsize=1.15\hsize\linewidth=\hsize\raggedright\arraybackslash}X>{\hsize=1.55\hsize\linewidth=\hsize\raggedright\arraybackslash}X>{\hsize=1.3\hsize\linewidth=\hsize\raggedright\arraybackslash}X>{\hsize=0.75\hsize\linewidth=\hsize\raggedright\arraybackslash}X>{\hsize=0.75\hsize\linewidth=\hsize\raggedright\arraybackslash}X>{\hsize=0.75\hsize\linewidth=\hsize\raggedright\arraybackslash}X>{\hsize=0.75\hsize\linewidth=\hsize\raggedright\arraybackslash}X@{}}
\toprule
Config & Provenance graph & Decomp./refinement guidance & Attribution input & $G_1$ & $G_2$ & $G_3$ & $G_4$ \\
\midrule
$M_{\mathrm{full}}$ & Yes & Yes & Dependency graph & \textbf{42.1\%} & \textbf{58.7\%} & \textbf{61.9\%} & \textbf{64.1\%} \\
$M_{\mathrm{guide}}$ & — & Yes & Execution trace & 33.9\% & 31.3\% & 18.8\% & — \\
$M_{\mathrm{trace}}$ & — & — & Execution trace & 37.6\% & 38.7\% & 40.0\% & 24.2\% \\
$M_{\mathrm{miss}}$ & — & — & Misses only & 30.0\% & 26.4\% & 24.5\% & 22.1\% \\
\bottomrule
\end{tabularx}
\end{table*}

\textbf{(2) Version adoption across eight rounds.} Three rounds adopt every edit and five remove some, lifting training hits from 64 to 182 (Table~\ref{tab:5}, Appendix Figure~\ref{fig:8}).

\begin{table*}[t]
\centering\footnotesize\setlength{\tabcolsep}{7pt}\renewcommand{\arraystretch}{0.94}
\caption{End-to-end progression across all eight evolution rounds ($G_0\to G_8$), separating pooled C/H/M total hits (across all 79 GTs $\times$ 3 runs = 237) and their net gain from equal-weighted per-project average recall (matching Figure~\ref{fig:5}), its net gain, and adoption decisions.}\label{tab:6.4}\label{tab:5}
\begin{tabular}{@{}llllll@{}}
\toprule
Round & Total C/H/M Hits (/237) & Hit $\Delta$ & Per-Proj Avg Recall & Recall $\Delta$ & Decision \\
\midrule
$G_0 \to G_1$ & 64 $\to$ 92 & +28 & 28.7\% $\to$ 42.1\% & +13.4\% & Partial adoption (3/7 kept) \\
$G_1 \to G_2$ & 92 $\to$ 116 & +24 & 42.1\% $\to$ 58.7\% & +16.6\% & Full adoption (3/3 kept) \\
$G_2 \to G_3$ & 116 $\to$ 136 & +20 & 58.7\% $\to$ 61.9\% & +3.2\% & Partial adoption (2/3 kept) \\
$G_3 \to G_4$ & 136 $\to$ 139 & +3 & 61.9\% $\to$ 64.1\% & +2.2\% & Full adoption (2/2 kept) \\
$G_4 \to G_5$ & 139 $\to$ 147 & +8 & 64.1\% $\to$ 68.5\% & +4.4\% & Full adoption (3/3 kept) \\
$G_5 \to G_6$ & 147 $\to$ 153 & +6 & 68.5\% $\to$ 71.6\% & +3.1\% & Partial adoption (2/6 kept) \\
$G_6 \to G_7$ & 153 $\to$ 173 & +20 & 71.6\% $\to$ 77.6\% & +6.0\% & Partial adoption (2/8 kept) \\
$G_7 \to G_8$ & 173 $\to$ 182 & +9 & 77.6\% $\to$ 82.1\% & +4.5\% & Partial adoption (4/8 kept, final) \\
\bottomrule
\end{tabular}
\end{table*}

\begin{figure*}[t]
\centering
\includegraphics[width=\textwidth]{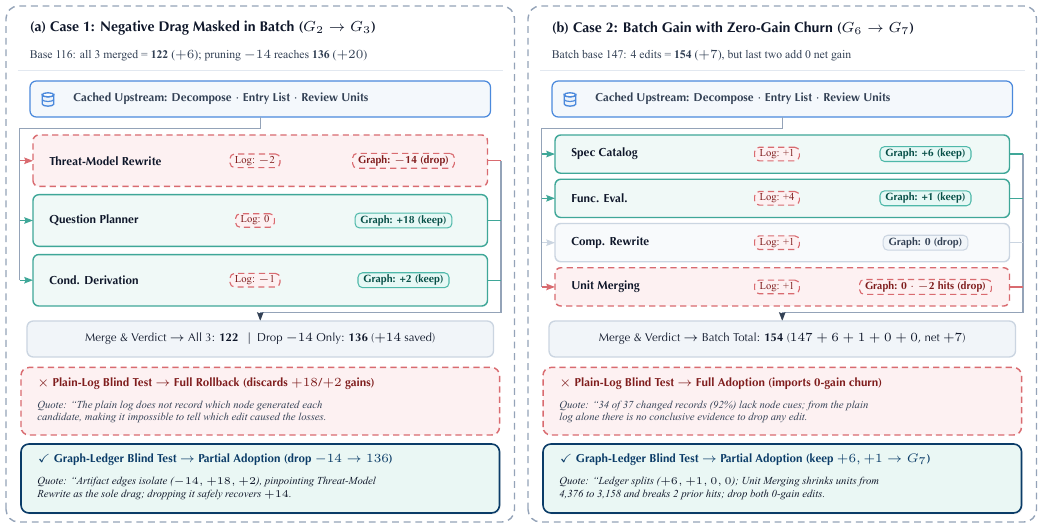}
\caption{Isolated blind-test comparison of Graph-Ledger Attribution vs. Plain-Log Attribution on two representative partial-adoption rounds: (a) $G_2\to G_3$ under negative batch drag and (b) $G_6\to G_7$ under mixed gain with zero-gain churn. Each panel shows per-node credit badges alongside verbatim quotes and adoption decisions of the two isolated analyst agents.}
\label{fig:9}
\end{figure*}

\textbf{(3) Plain-log versus graph decisions.} We package $G_2\to G_3$ and $G_6\to G_7$ twice each: a plain-log bundle with final finding text, total scores, and execution traces (the same attribution input as $M_{\mathrm{guide}}$), and a graph bundle that adds artifact read-write edges and the node provenance ledger. Two isolated, identically configured analyst agents that differ only in this input decide independently (Figure~\ref{fig:9}, Table~\ref{tab:11}). The plain-log agent rolled back everything on $G_3$ and adopted everything on $G_7$; the graph agent adopted partially by the ledger, and both re-runs support its decisions.

\begin{table*}[t]
\centering\small
\caption{Decisions of the plain-log and graph analyst agents on $G_3$ and $G_7$ (training C/H/M total hits out of 237; the $G_7$ batch exam uses that round's batch judge baseline $G_6=147$, while the adopted column reports the unified 10-project × 3-run evaluation).}\label{tab:6.5}\label{tab:11}
\begin{tabularx}{\textwidth}{@{}l>{\hsize=0.55\hsize\linewidth=\hsize\raggedright\arraybackslash}X>{\hsize=0.55\hsize\linewidth=\hsize\raggedright\arraybackslash}X>{\hsize=2.1\hsize\linewidth=\hsize\raggedright\arraybackslash}X>{\hsize=0.8\hsize\linewidth=\hsize\raggedright\arraybackslash}X@{}}
\toprule
Round & All merged & Plain-log decision & Graph decision (ledger net gain) & Adopted result (10 projects × 3 runs) \\
\midrule
$G_2\to G_3$ & 116 $\to$ 122 & Roll back all & Drop threat-model rewrite (−14); keep question planner (+18) and condition derivation (+2) & 122 $\to$ \textbf{136} \\
$G_6\to G_7$ & 147 $\to$ 154 & Adopt all & Drop comparison rewrite and unit merging (zero gain, broke stable hits); keep catalog modeling (+6) and function evaluation (+1) & 154 $\to$ \textbf{173} \\
\bottomrule
\end{tabularx}
\end{table*}

\textbf{Summary (RQ3).} The full configuration keeps climbing in early rounds while the ablations fall back; ledger-guided removal lifts $G_3$ from 122 to 136 hits and, on $G_7$, dropping two zero-gain edits reaches 173 hits with all damaged stable hits restored.

\subsection{RQ4: Real-world audit contest evaluation}\label{sec:6.5}

\textbf{Question.} On a live real-world smart contract audit contest, how many organizer-confirmed vulnerabilities does the evolved LucidScan workflow report, and how does it rank?

\textbf{Results.} Beyond offline benchmarks, we deployed the evolved workflow in an anonymized live public audit contest concluded in November 2026 (896 submitting researchers, USD 50,000 prize pool, 72 confirmed scored H/M clusters, and 452 confirmed clusters across all severities; project and platform names withheld). Because LucidScan is an upstream discovery workflow, live submissions paired it with scope filtering against contest rules, deduplication, and proof-of-concept verification. Out of 84 submitted findings, 78 were confirmed valid (92.9\% validity, 6 rejected), covering 74 distinct clusters (\#2 overall). As Table~\ref{tab:12} shows, the workflow found 6 of 8 confirmed High-severity issues (75.0\%, tied \#1 overall) and 28 of 64 Medium-severity issues (43.8\%, tied \#2 overall), totaling 34 of 72 scored H/M clusters (47.2\%, \#2 overall), and finished \#3 overall on the final leaderboard (7.24 points, USD 1,136).

\begin{table}[!tb]
\centering\footnotesize\setlength{\tabcolsep}{3.5pt}\renewcommand{\arraystretch}{1.04}
\caption{Results on an anonymized live public audit contest concluded in November 2026 (896 submitting researchers, USD 50,000 prize pool; project and platform names withheld; \#3 overall leaderboard finish with 7.24 points).}\label{tab:6.6}\label{tab:12}
\begin{tabularx}{\columnwidth}{@{}l>{\hsize=0.78\hsize\linewidth=\hsize\raggedright\arraybackslash}X>{\hsize=0.78\hsize\linewidth=\hsize\raggedright\arraybackslash}X>{\hsize=0.9\hsize\linewidth=\hsize\raggedright\arraybackslash}X>{\hsize=1.54\hsize\linewidth=\hsize\raggedright\arraybackslash}X@{}}
\toprule
Severity & Found & Total & Recall & Rank (896) \\
\midrule
High & 6 & 8 & 75.0\% & Tied \#1 \\
Medium & 28 & 64 & 43.8\% & Tied \#2 \\
\textbf{Scored H/M} & \textbf{34} & \textbf{72} & \textbf{47.2\%} & \textbf{\#2 overall} \\
Low & 24 & 151 & 15.9\% & \#3 overall \\
Info & 16 & 229 & 7.0\% & Tied \#6 \\
\textbf{All confirmed clusters} & \textbf{74} & \textbf{452} & \textbf{16.4\%} & \textbf{\#2 overall} \\
\bottomrule
\end{tabularx}
\end{table}

\textbf{Anonymized case study.} One confirmed High-severity finding (reported by only 13 of 896 researchers) involved an automated range-rebalancing hook attached to a lending-backed liquidity pool. Because the hook allowed repeated rebalancing triggers within a single transaction without bounding cumulative state drift, any unprivileged swapper could repeatedly cross the trigger range and drain 14\%–34\% of LP position value within 10 transactions, as verified by a mainnet-fork PoC. Surfacing this flaw required tracking multi-step state updates across the swap callback and rebalancing module while checking repeated boundary crossings, matching the cross-unit interaction and numeric boundary steps evolved into $G_8$.

\textbf{Summary (RQ4).} On an anonymized live public audit contest with 896 researchers concluded in November 2026, the evolved workflow tied for \#1 in High-severity recall (6/8, 75.0\%), ranked \#2 in scored H/M clusters (34/72, 47.2\%) with 92.9\% validity (78/84), and finished \#3 overall.

\section{Related Work}\label{sec:7}
Our work relates to three lines of research: LLM-based smart contract auditing, agent self-construction and workflow evolution, and runtime security-agent adaptation.

\textbf{LLM-based smart contract auditing.} GPTScan ~\cite{sun2024gptscan} combines LLMs with static analysis to match predefined vulnerability types. PropertyGPT ~\cite{liu2025propertygpt} uses source retrieval to guide LLMs in generating security properties for formal verification. KnowDit ~\cite{kong2026knowdit} builds a knowledge graph from historical audit reports to generate project-specific audit specifications and PoC test scaffolds. Rather than attaching a historical vulnerability knowledge base or generating project-specific specifications, Shuttle evolves the structure and general review steps of the upstream discovery workflow offline, then evaluates the frozen workflow by ground-truth root-cause recall.

\textbf{Agent self-construction and workflow evolution.} ADAS ~\cite{hu2025adas} and DGM ~\cite{zhang2025dgm} search agent implementations in code space, while AFlow ~\cite{zhang2025aflow} optimizes workflows with Monte Carlo tree search over end-to-end execution scores. Harness$^2$ ~\cite{xu2026harness2} runs a “Propose–Probe–Compose” loop from end-to-end task evaluations, and AgentFlow ~\cite{liu2026agentflow} diagnoses and rewrites a typed agent graph from target runtime feedback. In vulnerability detection, EvoHunt ~\cite{wang2026evohunt} and MetaVul ~\cite{ceka2026metavul} evolve workflows and procedural knowledge via multi-agent reflection or paired vulnerable-and-patched function feedback. These methods update workflows primarily from end-to-end scores or full run logs. Shuttle instead expands an instance-level artifact read-write dependency graph, generates proposals from stage-level bottlenecks with project source code and GTs isolated, and filters merged edits using a node-level provenance ledger and local re-runs.

\textbf{Security-agent evolution and constrained self-improvement.} CyberEvolver ~\cite{cyberevolver} runs per-target population evolution at execution time for CTF and penetration testing. RRSI ~\cite{rrsi} curbs overfitting in self-improvement through proposal constraints, case-specific screening, and structural pruning. Shuttle targets cross-project smart contract auditing, combining generality blind review with bidirectional attribution and protected-set verification on the run dependency graph.

\section{Discussion and Limitations}\label{sec:8}
Our evaluation has three main limitations:

\begin{enumerate}
\item \textbf{LLM judge.} Attribution uses \texttt{Sol medium} and root-cause matching relies on automatic \texttt{Sol high} judgments. Section 6.1 reports an independent heterogeneous-model audit of 60 stratified root-cause judgments against contract source code (58/60 agreement, $\kappa=0.933$); larger-scale blind annotation by human security experts remains future work.
\item \textbf{Single full evolution trajectory.} Three runs of each frozen version measure single-audit sampling variance. Because of the API cost of a full evolution cycle, we ran one complete mainline trajectory from $G_0$ to $G_8$.
\item \textbf{Downstream filtering.} Downstream severity filtering, cross-report clustering, and automated PoC validation are outside this paper's scope and remain future work.
\end{enumerate}

\section{Conclusion}\label{sec:9}
We presented Shuttle, a constrained workflow-evolution system for smart contract auditing. Through provenance-graph attribution, Shuttle locates where candidate clues are lost, traces where existing hits originate, and filters merged edits with a node provenance ledger while generating general workflow revisions with project source code and ground truth isolated. Our evaluation supports four conclusions. First, under the same base model on 13 validation projects not used in $M_{\mathrm{full}}$ evolution, the evolved LucidScan $G_8$ raises per-project average C/H/M recall from 28.6\% at $G_0$ to 70.3\%, above Pashov and Codex Security under their default settings, at USD 4.92 direct API cost per validation project per run. Second, the weakly constrained pilot $M_{\mathrm{weak}}$ reaches 88.5\% development recall without generality blind review, but incurs a 26.6× inflation in per-project scan tokens and a 31\%–96\% validation divergence. Third, in the early-round ablation window, the full configuration keeps climbing while all three graph-free ablations regress; on $G_3$ and $G_7$, removing negative or stability-harming edits by the ledger restores total hits or stable hits. Fourth, on an anonymized live public audit contest with 896 researchers concluded in November 2026, the evolved scanner tied for \#1 overall in High-severity recall (6/8, 75.0\%), ranked \#2 overall in scored H/M clusters (34/72, 47.2\%) with 92.9\% submission validity (78/84), and finished \#3 overall.

\bibliographystyle{ACM-Reference-Format}
\bibliography{references}
\appendix
\section{Reproducibility Details}\label{sec:A}
\subsection{Edit scope and reproducibility artifacts}\label{sec:A.1}

Node-content edits affect prompt fragments, fields, and output contracts. Graph edits add or remove nodes, connections, or execution policies. All edits require dependency, schema, and final-output compatibility checks.

Each round retains application snapshots, proposal inputs, candidate edits, reviews, merge manifests, scan records, and blind-label mappings. Stage-level process bottleneck summaries and the experience log are tied to concrete versions. Final $G_8$ incorporates a cross-unit relation matrix, concrete boundary evaluation, documentation-to-function expectation alignment, and explicit forwarding of key expressions and extreme values.

\subsection{Validation projects and repetitions}\label{sec:A.2}

\begin{figure}[!tb]
\centering
\includegraphics[width=\columnwidth]{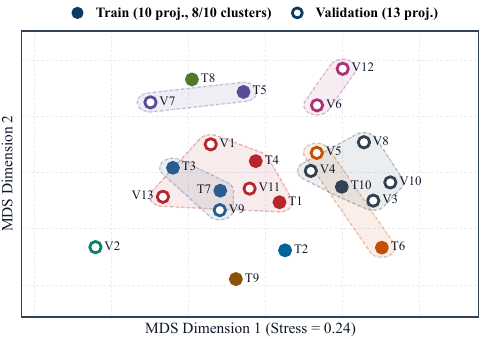}
\caption{Affinity Propagation clustering of the 23 benchmark projects (311 deduplicated contracts and libraries, 10 clusters). Filled circles denote the 10 training projects, hollow circles denote the 13 validation projects, colors distinguish clusters, and shaded dashed hulls group multi-project clusters.}
\label{fig:4}
\end{figure}

The 13 validation projects are anonymized as \texttt{V1}–\texttt{V13} (and the 10 training projects as \texttt{T1}–\texttt{T10}). Figure~\ref{fig:4} visualizes the 2D Affinity Propagation clustering of the 23 benchmark projects (311 deduplicated contracts and libraries). On validation, both paired baseline $G_0$ and final $G_8$ run three passes as main validation results, while $G_6$ and $G_7$ run two passes each (Figure~\ref{fig:5} plots the paired $G_0$ baseline and matched two-pass validation trajectory alongside the full three-pass $G_8$ outcome). The main training comparison uses three runs per version. Critical/high/medium denominators are 79 per training run and 114 per validation run.

\subsection{Detailed $M_{\mathrm{full}}$ evolution rounds}\label{sec:A.3}

\begin{figure*}[t]
\centering
\includegraphics[width=\textwidth]{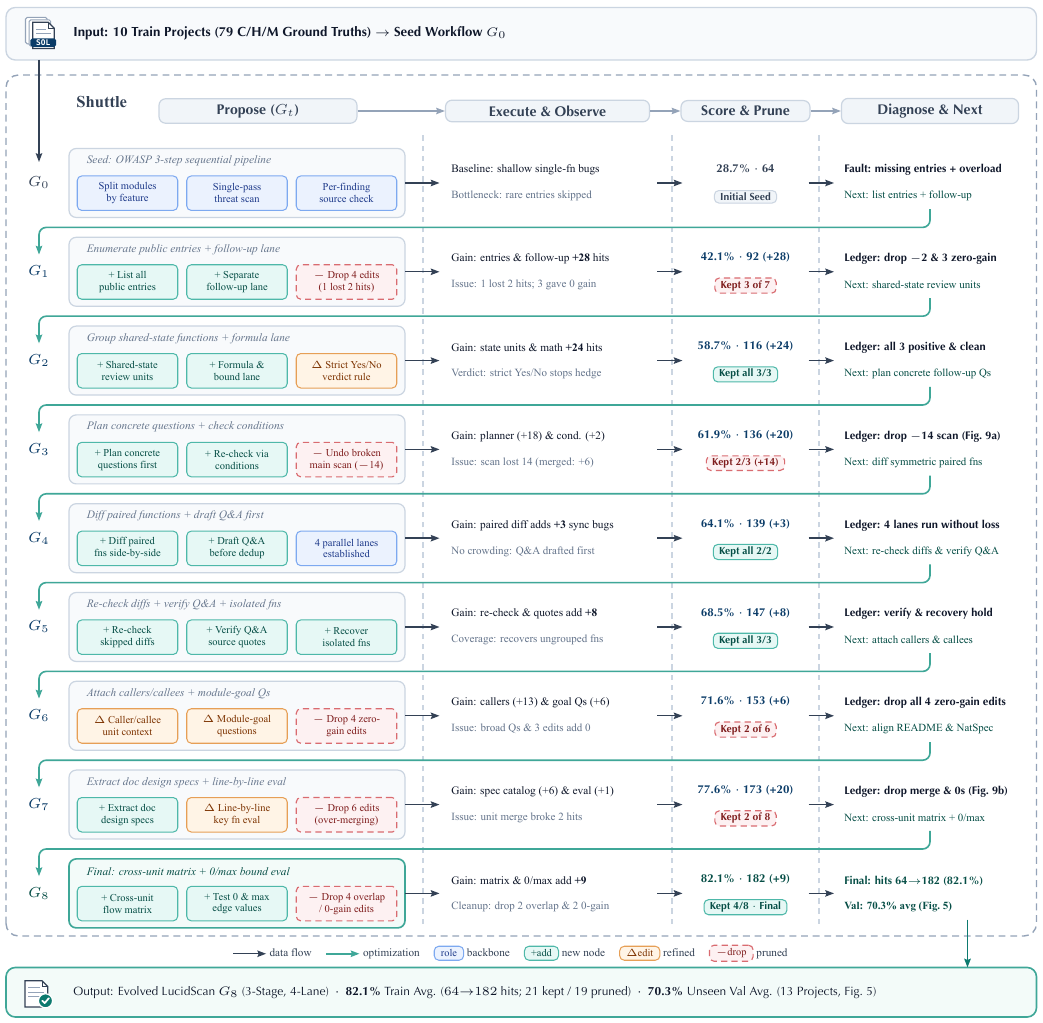}
\caption{Shuttle's complete eight-round constrained evolution matrix from $G_0$ to $G_8$ with no skipped rounds (all hit counts and recall figures match Table~\ref{tab:5}). Each row aligns workflow topology operators (blue backbone, \texttt{+} teal added, \texttt{Δ} ochre edited, \texttt{−} crimson-dashed dropped), observed runtime gains and bottlenecks, training recall and retention ratio, and dependency-graph attribution feeding into the next round.}
\label{fig:8}
\end{figure*}

Figure~\ref{fig:8} shows the full eight-round evolution loop from $G_0$ to $G_8$, Table~\ref{tab:6} lists every $M_{\mathrm{full}}$ evolution round from $G_0$ through final $G_8$, and the inline table below gives the complete selection record of the candidate versions for $G_6 \to G_7$ in recall and protected-set stability.

\begin{table*}[t]
\centering\footnotesize\renewcommand{\arraystretch}{1.05}
\caption{Detailed $M_{\mathrm{full}}$ evolution rounds ($G_0$ through $G_8$): target stages, design changes, proposal scale, and adoption outcomes (corresponding to Figure~\ref{fig:5}).}\label{tab:A.0}\label{tab:6}
\begin{tabularx}{\textwidth}{@{}l>{\hsize=0.65\hsize\linewidth=\hsize\raggedright\arraybackslash}X>{\hsize=1.35\hsize\linewidth=\hsize\raggedright\arraybackslash}X>{\hsize=1.0\hsize\linewidth=\hsize\raggedright\arraybackslash}X>{\hsize=1.0\hsize\linewidth=\hsize\raggedright\arraybackslash}X@{}}
\toprule
Round / Version & Target Stage & Design Idea (Plain Language) & Proposals \& Scale & Outcome \\
\midrule
$G_0$ & Full 3-step baseline & Basic OWASP flow: split modules, model threats, then reason & 3 base nodes & Starting baseline (28.7\% proj avg) \\
$G_1$ & Decomp. \& Threat Modeling & List entry points first before hunting bugs; add a follow-up Q\&A lane & 7 proposed, 3 kept (+2 nodes, 1 prompt) & Kept 3 after cached leave-one-out pruning of 4 redundant edits \\
$G_2$ & All 3 stages & Group related functions into review units and add formula derivation & 3 proposed, 3 kept (+1 node, unit/verdict) & Adopted (+30.0 pp over $G_0$ cumulative) \\
$G_3$ & Threat Modeling & Plan supplementary questions explicitly and solve conditions step-by-step & 3 proposed, 2 kept (+2 nodes, 1 pruned) & Adopted after pruning a harmful modeling rewrite \\
$G_4$ & Threat Modeling & Compare related code locations side-by-side; draft \& dedup independently & 2 proposed, 2 kept (+1 node, dedup edit) & Adopted; small steady gain \\
$G_5$ & Threat Modeling & Double-check comparisons and verify supplementary answers against questions & 3 proposed, 3 kept (+2 nodes, 1 scope) & Adopted (+4.4 pp over $G_4$) \\
$G_6$ & Decomp. \& Modeling & Build on ungrouped-function recovery with call-endpoint pairing and expectation-guided Q\&A & 9 proposed, 6 evaluated, 2 kept (+2 edits) & Adopted (train avg 71.6\%, union 75.9\%) \\
$G_7$ & Decomp. \& Modeling & Build a project spec/reference catalog and evaluate statements around key functions & 12 proposed, 8 approved in two 4-edit batches, 2 kept (+1 node, 1 prompt) & Dropped −2 batch \& 2 zero-gain edits from +7 batch; re-run (10 projects × 3 runs) reached 77.6\% and restored all broken stable hits \\
$G_8$ & All 3 stages & Add relation matrix, concrete boundary substitution, doc-expectation alignment \& boundary parameter passing & 8 approved, 4 kept (+2 nodes, 2 handoff/list) & Final version ($G_8$, 82.1\% train avg / 70.3\% val avg) \\
\bottomrule
\end{tabularx}
\end{table*}

\par\smallskip\noindent{\footnotesize\setlength{\tabcolsep}{2.8pt}\renewcommand{\arraystretch}{1.1}\begin{tabularx}{\columnwidth}{@{}>{\raggedright\arraybackslash}X>{\raggedright\arraybackslash}X>{\raggedright\arraybackslash}X>{\raggedright\arraybackslash}X@{}}
\toprule
Candidate ($G_6 \to G_7$) & C/H/M Hits (/237) & Protected Set $\Phi_t$ Stability & Rule Decision \\
\midrule
Batch baseline $G_6$ & 147 & $\ge 2$-pass: 49; 3/3 stable: 38 & Round-7 batch baseline \\
Alt. 4-edit batch & 145 (−2) & Stable hits regressed & \textbf{Reject batch} (negative gain) \\
Figure~\ref{fig:9}(B) 4-edit batch & 154 (+7) & 3/3 stable hits unchanged (2 dropped 3$\to$2; 1 crit. 2$\to$0) & \textbf{Prune} (drop 2 zero-gain, stability-harming edits) \\
\textbf{Pruned $G_7$ (2 kept)} & \textbf{173 (77.6\%)} & \textbf{Both 3/3 restored; crit. restored 2/3} & \textbf{Adopt as $G_7$} \\
\bottomrule
\end{tabularx}}\par\smallskip

\subsection{Per-project scan cost breakdown of final $G_8$}\label{sec:A.4}

Table~\ref{tab:7} lists the per-project mean direct API scan cost of final $G_8$ across all 23 benchmark projects (3 runs each, 69 scans total), priced at USD 0.10/M uncached input, USD 0.01/M cached input, and USD 0.50/M output tokens, excluding offline evolution.

\begin{table*}[t]
\centering\small
\caption{Per-project mean scan cost of $G_8$ across all 23 benchmark projects (3 runs each, 69 scans total; USD derived from recorded run token logs including retries, excluding offline evolution).}\label{tab:A.1}\label{tab:7}
\begin{tabularx}{\textwidth}{@{}l>{\raggedright\arraybackslash}X>{\raggedright\arraybackslash}X>{\raggedright\arraybackslash}X>{\raggedright\arraybackslash}X>{\raggedright\arraybackslash}X@{}}
\toprule
Train Project (3r) & Mean Cost (USD) & Val Project (3r) & Mean Cost (USD) & Val Project (cont.) & Mean Cost (USD) \\
\midrule
T1 & 3.42 & V1 & 3.67 & V8 & 3.02 \\
T2 & 1.97 & V2 & 0.97 & V9 & 4.18 \\
T3 & 1.25 & V3 & 7.46 & V10 & 4.68 \\
T4 & 5.09 & V4 & 7.10 & V11 & 6.01 \\
T5 & 2.95 & V5 & 5.48 & V12 & 3.68 \\
T6 & 3.82 & V6 & 11.71 & V13 & 2.98 \\
T7 & 3.83 & V7 & 3.08 & — & — \\
T8 & 1.18 & \textbf{Val Avg (39 scans)} & \textbf{4.92 (192.1 total)} & \textbf{Overall (69 scans)} & \textbf{3.97 (273.9 total)} \\
T9 & 1.31 &  &  &  &  \\
T10 & 2.45 &  &  &  &  \\
\textbf{Train Avg (30 scans)} & \textbf{2.73 (81.8 total)} &  &  &  &  \\
\bottomrule
\end{tabularx}
\end{table*}

\end{document}